\documentclass[10pt,journal]{IEEEtran}

\ifx\pdfoutput\undefined
\usepackage{graphicx}
\graphicspath{{eps_figs/}}
\else
\usepackage[pdftex]{graphicx}
\graphicspath{{pdf_figs/}}
\fi

\usepackage{latexsym,amssymb,amsmath,cite,ifthen,afterpage}
\usepackage{setspace}

\newcommand{\Fig}[1]{Fig.~\ref{#1}}

\newcommand{\eqdef}{\stackrel{\scriptscriptstyle\bigtriangleup}{=} }

\newcommand{\calX}{\mathcal{X}}
\newcommand{\calS}{\mathcal{S}}

\newcommand{\x}{\textbf{x}}
\newcommand{\X}{\textbf{X}}
\newcommand{\y}{\textbf{y}}
\newcommand{\Y}{\textbf{Y}}

\newcommand{\E}{\operatorname{E}}

\newcounter{examplecntr}
{\begin{trivlist}\small\item[]\refstepcounter{examplecntr}%
 {\bfseries Example~\theexamplecntr%
  \ifthenelse{\equal{#1}{}}{}{ (#1)}
}}%
{\end{trivlist}}

\newcounter{theoremcntr}
{\begin{trivlist}\item[]\refstepcounter{theoremcntr}%
{\bfseries Theorem~\thetheoremcntr%
  \ifthenelse{\equal{#1}{}}{}{ (#1)}.
}}%
{\hfill$\Box$\end{trivlist}}

\newcommand{\cent}[1]{\makebox(0,0){#1}}
\newcommand{\pos}[2]{\makebox(0,0)[#1]{#2}}

\begin{document}
\DeclareGraphicsExtensions{.pdf}

\title{Generalized Belief Propagation for the\\ Noiseless Capacity and Information Rates of Run-Length Limited
Constraints}

\author{Giovanni Sabato,~\IEEEmembership{Member,~IEEE} and Mehdi Molkaraie,~\IEEEmembership{Member,~IEEE}

\thanks{%
Giovanni Sabato 
is with PARALLEL Informatik AG, CH-6005 Luzern, Switzerland. 
Mehdi Molkaraie is with the Dept.\ of Information Technology and Electrical Engineering, 
ETH Z\"urich, CH-8092 Z\"urich, Switzerland. 
Emails: giovanni.sabato@parallel.ch, molkaraie@isi.ee.ethz.ch.
}}



\maketitle

\begin{abstract}
The performance of the generalized belief propagation algorithm 
to compute the noiseless capacity and mutual \mbox{information} rates of 
finite-size two-dimensional and three-dimensional run-length limited 
constraints is investigated. In both cases, the problem is reduced to
estimating the partition function of graphical models with cycles. The
partition function is then estimated using the region-based free energy 
approximation technique. For each constraint, a method is 
proposed to choose the basic regions and to construct the region graph
which provides the graphical framework
to run the generalized belief propagation algorithm.
Simulation results for the noiseless capacity of different constraints as a 
function of the size of the channel are reported.
In the cases that tight lower and upper bounds on the Shannon capacity exist,
convergence to the Shannon capacity is discussed.
For noisy constrained channels, 
simulation results are reported for
mutual information rates as a function of signal-to-noise ratio.
\end{abstract}


\section{Introduction}
\label{sec:Introduction}

Run-length limited (RLL) constraints are widely used in magnetic and optical recording
systems. Such constraints reduce the effect of inter-symbol interference and help
in timing control. In track-oriented storage systems constraints are defined
in one dimension. 

We say a binary one-dimensional (1-D) sequence satisfies the 
$(d,k)$-RLL constraint if the 
runs of $0$'s have length at most $k$ and the runs of $0$'s between 
successive $1$'s have length at least $d$. 
We suppose that $0 \le d < k \le \infty$. 

The {\it Shannon capacity} of a 1-D $(d,k)$-RLL constraint is defined as
\begin{equation} \label{eqn:Cap1}
C_{1D}^{(d,k)} \eqdef \lim_{m \to \infty} \frac{\log_2 Z(m)}{m},
\end{equation}
where $Z(m)$ denotes the number of binary 1-D sequences of length $m$ that 
satisfy the $(d,k)$-RLL constraint, see~\cite{Sch:04, Sh:48}.

With the rise in demand for larger storage in smaller size and with recent developments 
in page-oriented storage systems, such as holographic
data storage,
two-dimensional (2-D) 
constraints have become more of interest \cite{Sie:06}. In 
these systems, data is organized on a surface and constraints are defined 
in two dimensions.

A 2-D binary array satisfies the $(d_1,k_1,d_2,k_2)$-RLL constraint if it
satisfies a $(d_1,k_1)$-RLL constraint horizontally and a $(d_2,k_2)$-RLL
constraint vertically. If a 2-D binary array satisfies a 1-D $(d,k)$-RLL 
constraint both horizontally and vertically, we simply say that it 
satisfies a 2-D $(d,k)$-RLL constraint.

\emph{Example: 2-D $(2,\infty)$-RLL constraint:}

The 2-D $(2,\infty)$-RLL constraint is satisfied in the following 2-D 
binary array segment. In words, in every row and every column of the 
array there are at least two $0$'s between successive $1$'s; but the 
runs of $0$'s can be of any length (however, $1$'s can be 
diagonally adjacent).

\begin{center}

$\ldots0100100001001000100000100010\ldots$\linebreak
$\ldots1000010000100010000100000100\ldots$\linebreak
$\ldots0001000010000001000000010001\ldots$\linebreak
$\ldots0100100100010000001000100000\ldots$\linebreak
\end{center}

The Shannon capacity of a 2-D $(d_1,k_1,d_2,k_2)$-RLL constraint 
is defined as
\begin{equation} 
\label{Cap2}
C_{2D}^{(d_1,k_1,d_2,k_2)} \eqdef \lim_{m,n \to \infty} \frac{\log_2 Z(m,n)}{mn},
\end{equation}
where $Z(m,n)$ denotes the number of 2-D binary arrays of size $m\times n$ 
that satisfy the $(d_1,k_1,d_2,k_2)$-RLL constraint.

Similarly, the Shannon capacity can be defined for higher dimensional 
constrained channels. For example, the Shannon capacity in three dimensions 
$C_{3D}^{(d_1,k_1,d_2,k_2,d_3,k_3)}$
depends on $Z(m,n,q)$, the number of three-dimensional (3-D) 
binary arrays of size $m\times n\times q$
that satisfy a $(d_1,k_1,d_2,k_2,d_3,k_3)$-RLL constraint.

The noiseless capacity of a constrained channel is an important quantity 
that provides an upper bound to the 
information rate of any encoder that maps arbitrary binary input into binary 
data that satisfies a given constraint.
There are a number of techniques to compute the \mbox{1-D} Shannon capacity 
(for example combinatorial or algebraic approaches) \cite{Sch:04}. In contrast 
to the 1-D capacity, except for a few cases, exact values 
of two and higher dimensional (positive) Shannon capacities are 
not known, see~\cite{CW:98, FL:07, KZ:99, IKNZ:00, OR:09, TR:09}.

For noisy \mbox{1-D} constrained channels, simulation-based techniques proposed 
in~\cite{ALVKZ:sbcir2006, PSS:2001} can be 
used to compute mutual
information rates. However, computing 
mutual information rates of noisy \mbox{2-D} RLL constraints has been an unsolved 
problem.

In this paper, the goal is to apply the 
generalized belief propagation (GBP) algorithm~\cite{YFW:05}
for the above-mentioned problems, namely,
to compute an estimate of the capacity of 
noiseless \mbox{2-D} and \mbox{3-D} RLL constrained channels
and mutual information rates of noisy \mbox{2-D} constrained channels.
For both
problems GBP turns out to yield very good approximate results.

Preliminary versions of the material of this paper have appeared 
in~\cite{SM:10} and~\cite{ML:10}.
In~\cite{SM:10}, we applied GBP to compute the noiseless
capacity of \mbox{2-D} and \mbox{3-D} RLL constrained channels.
In~\cite{ML:10}, GBP was applied to compute mutual information rates 
of a \mbox{2-D} $(1,\infty)$-RLL constrained channel with relatively small size 
and only at high signal-to-noise ratio (SNR).
In this paper, we show that both problems reduce to estimating the partition 
function of graphical models with cycles. We then apply GBP to both problems
and consider new constraints and larger sizes of grid.

Our main motivations for this research were the successful 
application of GBP for information rates of
\mbox{2-D} finite-state channels with memory 
in~\cite{SSSKWW:2DChannels2008}, Kikuchi approximation for decoding of
LDPC codes and partial-response channels in~\cite{PA:06},
and tree-based 
Gibbs sampling for the noiseless capacity and information rates of 
\mbox{2-D} constrained channels in~\cite{LM:09, ML:10}.

The outline of the paper is as follows.
In Section~\ref{sec:Set-up}, we consider the problem of computing
the partition function and discuss how this problem 
is related to computing the noiseless capacity and 
information rates of constrained channels.
Region graphs, 
GBP, and region-based free energy are outlined in Section~\ref{sec:GBP}.
Section~\ref{sec:2Dexample} discusses the capacity of noiseless \mbox{2-D} 
constraints.
Numerical values and simulation results for the capacity of 
noiseless \mbox{2-D} and \mbox{3-D}
RLL constraints are reported in Section~\ref{sec:NumericsC}.
In Section~\ref{sec:IR}, we apply GBP to compute \mbox{mutual}
information rates of noisy \mbox{2-D} RLL constraints and report
numerical experiments for mutual information rates in 
Section~\ref{sec:NumericsI}.

\section{Problem Set-up}
\label{sec:Set-up}

Consider a \mbox{2-D} channel of size 
$N = m\times m$ 
with a set of $\X = \{X_1,X_2, \ldots, X_N\}$ random variables.
Let $x_i$ denote a realization of $X_i$ 
and let $\x$ denote $\{x_1,x_2, \ldots, x_N\}$.
We assume that each $X_i$
takes values in a finite set $\calX_i$.
Also let $\calX$ be the Cartesian product 
$\calX \eqdef \calX_1 \times \calX_2 \times \ldots \times \calX_N$.

In constrained channels, not all sequences of symbols 
from the channel alphabet $\calX$ are admissible.
Let $\calS_\X \subset\calX$ be the set of 
admissible input sequences.
We define the indicator function 
\begin{equation} \label{eqn:indicatorfunction}
f(\x) \eqdef \left\{
\begin{array}{rl}
1, & \x \in \calS_\X \\
0, & \x \notin \calS_\X
\end{array} \right.
\end{equation}

The \emph{partition function}
$Z$ is defined as
\begin{equation}
\label{Z}
Z \eqdef \sum_{\x \in \calX} f(\x).
\end{equation}

With the above definitions, $Z = |\calS_\X|$
is the number of sequences that satisfy a given constraint.
Therefore, computing the capacity of constrained channels as 
expressed in~\eqref{Cap2},
is closely related to computing the partition function as
in~\eqref{Z}.

Also note that with the above definitions
\begin{equation}
\label{Px}
p(\x) = \frac{f(\x)}{Z}
\end{equation}
is a probability mass function on $\calX$.

For a noisy \mbox{2-D} channel, let $\X$ be the input and
\mbox{$\Y = \{Y_1,Y_2, \ldots, Y_N\}$}
be the output of the channel. The mutual information rate is
\begin{equation} \label{IRate}
\frac{1}{N}I(\X;\Y) = \frac{1}{N}\big(H(\Y) - H(\Y|\X)\big).
\end{equation}

Let us suppose that
$H(\Y|\X)$ 
is analytically available.
In this case, the problem of estimating 
the mutual information rate reduces to estimating 
the entropy of the channel output, which is
\begin{equation} \label{HY}
H(\Y) = -\E\big[\log p(\Y) \big].
\end{equation}

As
in~\cite{ALVKZ:sbcir2006},
we can approximate the expectation 
in (\ref{HY}) by drawing $L$ samples $\y^{(1)},\y^{(2)}, \ldots, \y^{(L)}$
according to $p(\y)$ and use the empirical average as
\begin{equation} \label{HYE}
H(\Y) \approx -\frac{1}{L}\sum_{\ell = 1}^{L} \log(p(\y^{(\ell)})).
\end{equation}

Therefore, the problem of estimating the mutual information rate
reduces to computing $p(\y^{(\ell)})$ for $\ell = 1,2, \ldots, L$.

We will compute $p(\y^{(\ell)})$ based on
\begin{equation} \label{PY1}
p(\y^{(\ell)}) = \sum_{\x \in \calX} p(\x)p(\y^{(\ell)} | \x),
\end{equation}
which for a fixed $\y^{(\ell)}$ has also the form~(\ref{Z}) and
therefore requires the computation of a partition function.

RLL constraints impose restrictions on the values of variables that 
can be verified locally.
For example, in a 2-D $(1,\infty)$-RLL constraint no two 
(horizontally or vertically) adjacent variables can both have 
the value~$1$. The indicator function 
of this constraint factors into a product of kernels of the form
\begin{equation} \label{NoAdjacentOnes}
\kappa_a(x_i,x_j) = \left\{ \begin{array}{ll}
     0, & \text{if $x_i = x_j = 1$} \\
     1, & \text{else,}
  \end{array} \right.
\end{equation}
with one such kernel for each adjacent pair $(x_i, x_j)$.

The factorization with kernels as in \eqref{NoAdjacentOnes} can be represented with a
graphical model. In this paper, we focus on graphical models
defined in terms of {\it Forney factor graphs}.
Fig.~\ref{fig:2DGrid} shows
the Forney factor graph of a 2-D $(1,\infty)$-RLL constraint
where the boxes labeled ``$=$'' are equality 
constraints \cite{Lg:ifg2004}.
(\Fig{fig:2DGrid} may also be viewed as a factor graph as in \cite{KFL:fg2000} 
where the boxes labeled ``$=$'' are the variable nodes).

In general, we suppose that the indicator function
$f(\x)$
of an RLL constraint
factors into a product of non-negative local kernels
each having some subset of $\x$ as arguments; i.e.\
\begin{equation}
\label{Factors}
f(\x) = \prod_{a} f_a(\x_a),
\end{equation}
where $\x_a$ is a subset of $\x$ and each kernel $f_a(\x_a)$ has 
elements of $\x_a$ as \mbox{arguments}.

In this case, the partition function in~\eqref{Z} can be written as
\begin{equation}
\label{Z2}
Z = \sum_{\x \in \calX} \prod_{a} f_a(\x_a).
\end{equation}

If the factorization in~\eqref{Factors} yields 
a cycle-free factor graph (with not too many states),
the sum in~\eqref{Z2}, or equivalently the sum in~\eqref{Z}, can be computed 
efficiently by the sum-product
message passing algorithm~\cite{KFL:fg2000}.
However, for the examples we study in this paper,
like the Forney factor graph of a 2-D $(1,\infty)$-RLL constraint in Fig.~\ref{fig:2DGrid}, 
factor graphs contain (many short) cycles.
In such cases computing $Z$
requires a sum with an exponential number of terms and therefore
we are interested in applying approximate methods. 

Due to the presence of many short cycles in the factor graph 
representation of 2-D and 3-D RLL constraints, 
loopy belief propagation 
often fails to converge. As a result, we apply GBP to 
estimate $Z$, which then leads
to estimating the noiseless capacity and mutual information 
rates of RLL constraints.

\newcommand{\drawgrid}{%
\begin{picture}(76,64)(0,0)
\small
%
\put(0,60){\framebox(4,4){$=$}}
\put(4,62){\line(1,0){8}}
\put(12,60){\framebox(4,4){}}
\put(16,62){\line(1,0){8}}
\put(24,60){\framebox(4,4){$=$}}
\put(28,62){\line(1,0){8}}
\put(36,60){\framebox(4,4){}}
\put(40,62){\line(1,0){8}}
\put(48,60){\framebox(4,4){$=$}}
\put(52,62){\line(1,0){8}}
\put(60,60){\framebox(4,4){}}
\put(64,62){\line(1,0){8}}
\put(72,60){\framebox(4,4){$=$}}
\put(2,54){\line(0,1){6}}
\put(0,50){\framebox(4,4){}}
\put(2,50){\line(0,-1){6}}
\put(26,54){\line(0,1){6}}
\put(24,50){\framebox(4,4){}}
\put(26,50){\line(0,-1){6}}
\put(50,54){\line(0,1){6}}
\put(48,50){\framebox(4,4){}}
\put(50,50){\line(0,-1){6}}
\put(74,54){\line(0,1){6}}
\put(72,50){\framebox(4,4){}}
\put(74,50){\line(0,-1){6}}
\put(0,40){\framebox(4,4){$=$}}
\put(4,42){\line(1,0){8}}
\put(12,40){\framebox(4,4){}}
\put(16,42){\line(1,0){8}}
\put(24,40){\framebox(4,4){$=$}}
\put(28,42){\line(1,0){8}}
\put(36,40){\framebox(4,4){}}
\put(40,42){\line(1,0){8}}
\put(48,40){\framebox(4,4){$=$}}
\put(52,42){\line(1,0){8}}
\put(60,40){\framebox(4,4){}}
\put(64,42){\line(1,0){8}}
\put(72,40){\framebox(4,4){$=$}}
\put(2,34){\line(0,1){6}}
\put(0,30){\framebox(4,4){}}
\put(2,30){\line(0,-1){6}}
\put(26,34){\line(0,1){6}}
\put(24,30){\framebox(4,4){}}
\put(26,30){\line(0,-1){6}}
\put(50,34){\line(0,1){6}}
\put(48,30){\framebox(4,4){}}
\put(50,30){\line(0,-1){6}}
\put(74,34){\line(0,1){6}}
\put(72,30){\framebox(4,4){}}
\put(74,30){\line(0,-1){6}}
\put(0,20){\framebox(4,4){$=$}}
\put(4,22){\line(1,0){8}}
\put(12,20){\framebox(4,4){}}
\put(16,22){\line(1,0){8}}
\put(24,20){\framebox(4,4){$=$}}
\put(28,22){\line(1,0){8}}
\put(36,20){\framebox(4,4){}}
\put(40,22){\line(1,0){8}}
\put(48,20){\framebox(4,4){$=$}}
\put(52,22){\line(1,0){8}}
\put(60,20){\framebox(4,4){}}
\put(64,22){\line(1,0){8}}
\put(72,20){\framebox(4,4){$=$}}
\put(2,14){\line(0,1){6}}
\put(0,10){\framebox(4,4){}}
\put(2,10){\line(0,-1){6}}
\put(26,14){\line(0,1){6}}
\put(24,10){\framebox(4,4){}}
\put(26,10){\line(0,-1){6}}
\put(50,14){\line(0,1){6}}
\put(48,10){\framebox(4,4){}}
\put(50,10){\line(0,-1){6}}
\put(74,14){\line(0,1){6}}
\put(72,10){\framebox(4,4){}}
\put(74,10){\line(0,-1){6}}
\put(0,0){\framebox(4,4){$=$}}
\put(4,2){\line(1,0){8}}
\put(12,0){\framebox(4,4){}}
\put(16,2){\line(1,0){8}}
\put(24,0){\framebox(4,4){$=$}}
\put(28,2){\line(1,0){8}}
\put(36,0){\framebox(4,4){}}
\put(40,2){\line(1,0){8}}
\put(48,0){\framebox(4,4){$=$}}
\put(52,2){\line(1,0){8}}
\put(60,0){\framebox(4,4){}}
\put(64,2){\line(1,0){8}}
\put(72,0){\framebox(4,4){$=$}}
\end{picture}
}

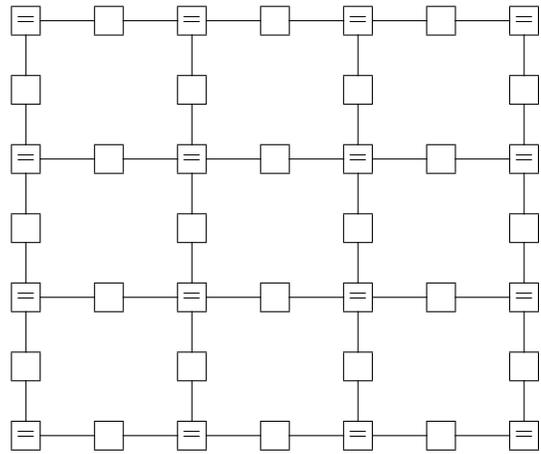
\begin{figure}[!t]
\setlength{\unitlength}{0.92mm}
\centering
\begin{picture}(76,64)(0,0)
\put(0,0){\drawgrid}
\centering
\end{picture}
\caption{\label{fig:2DGrid}
Forney factor graph for a 2-D $(1, \infty)$-RLL constraint. 
}
\end{figure}

\section{GBP and the region graph method}
\label{sec:GBP}

In statistical physics, $Z$ defined in~\eqref{Z} is 
known as the partition function and the 
\textit{Helmholtz free energy} is defined as
\begin{equation}
\label{FHZ}
F_H \eqdef -\ln(Z).
\end{equation}

The partition function and the Helmholtz free energy are  
important quantities in statistical physics since they carry 
information about all thermodynamic properties of a system.

A number of techniques have been developed in statistical physics 
to approximate the free energy.
The method that we apply in this paper is known as the region-based 
free energy approximation, in particular we use the cluster variation 
method to select a valid set of regions and counting 
numbers, see~\cite{YFW:05} and~\cite{Wel:04} for more details.

We start by introducing the region graph representation of our
problem. Such a region graph will provide a graphical framework for
GBP algorithm. For each RLL constraint, the size of the basic region is chosen
based on the constraint parameters. For a 2-D 
$(d_1,k_1,d_2,k_2)$-RLL constraint with finite $k_1$ and $k_2$, the
width and the height of the basic region is chosen as
\begin{eqnarray*}
\label{basicregionsize}
W_R & = & k_{1} + 1\\
H_R & = & k_{2}+ 1,
\end{eqnarray*}
and for the infinite case, the size is chosen as 
\begin{eqnarray*}
\label{basicregionsize2}
W_R & = & d_{1} + 1\\
H_R & = & d_{2}+ 1.
\end{eqnarray*}

\begin{figure}[t]
\centering
\includegraphics[width=6.05cm]{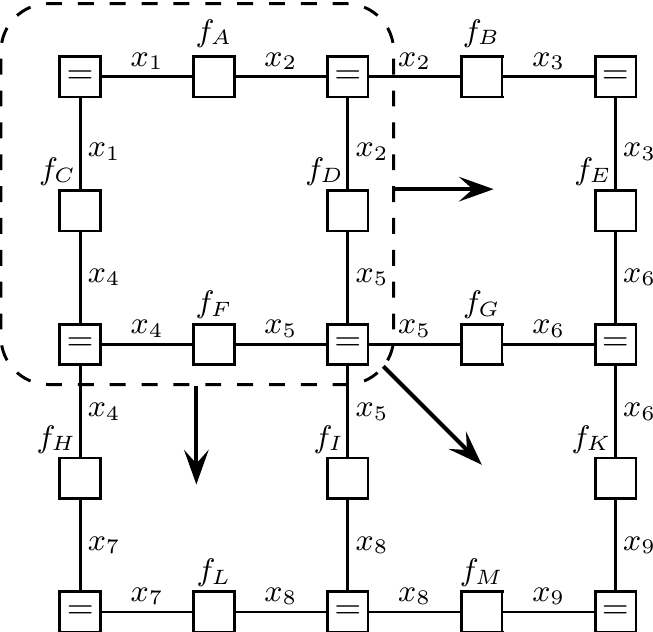}
\caption{\label{fig:rllfactorgraph}
Basic region of size $2 \times 2$ for a 2-D 
$(1, \infty)$-RLL constraint.}
\end{figure}

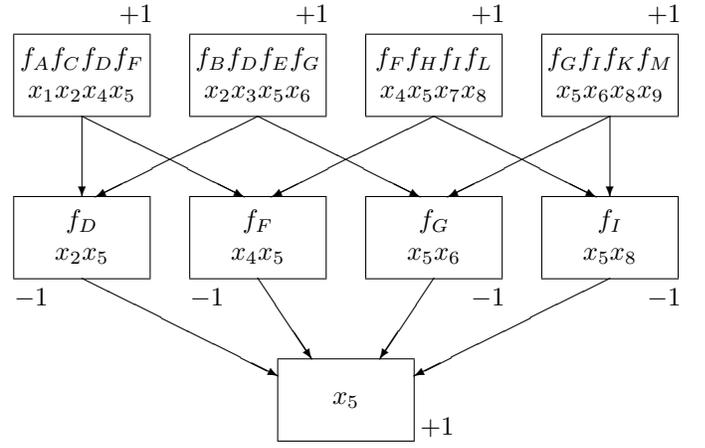
\begin{figure}
\centering
\setlength{\unitlength}{0.9mm}
\begin{picture}(98,65)(0,0)
%
\put(0,48){\framebox(20,12){$\begin{array}{c}f_A f_C f_D f_F \\ x_1 x_2 x_4 x_5 \end{array}$}}
 \put(18,63){\cent{$+1$}}
\put(26,48){\framebox(20,12){$\begin{array}{c}f_B f_D f_E f_G \\ x_2 x_3 x_5 x_6 \end{array}$}}
 \put(44,63){\cent{$+1$}}
\put(52,48){\framebox(20,12){$\begin{array}{c}f_F f_H f_I f_L \\ x_4 x_5 x_7 x_8 \end{array}$}}
 \put(70,63){\cent{$+1$}}
\put(78,48){\framebox(20,12){$\begin{array}{c}f_G f_I f_K f_M \\ x_5 x_6 x_8 x_9 \end{array}$}}
 \put(96,63){\cent{$+1$}}
\put(10,48){\vector(0,-1){12}}
\put(10,48){\vector(2,-1){24}}
\put(36,48){\vector(-2,-1){24}}
\put(36,48){\vector(2,-1){24}}
\put(62,48){\vector(-2,-1){24}}
\put(62,48){\vector(2,-1){24}}
\put(88,48){\vector(-2,-1){24}}
\put(88,48){\vector(0,-1){12}}
\put(0,24){\framebox(20,12){$\begin{array}{c}f_D \\ x_2 x_5 \end{array}$}}
 \put(2.5,21){\cent{$-1$}}
\put(26,24){\framebox(20,12){$\begin{array}{c}f_F \\ x_4 x_5 \end{array}$}}
 \put(28.5,21){\cent{$-1$}}
\put(52,24){\framebox(20,12){$\begin{array}{c}f_G \\ x_5 x_6 \end{array}$}}
 \put(70,21){\cent{$-1$}}
\put(78,24){\framebox(20,12){$\begin{array}{c}f_I \\ x_5 x_8 \end{array}$}}
 \put(96,21){\cent{$-1$}}
\put(10,24){\vector(2,-1){29}}
\put(36,24){\vector(2,-3){8}}
\put(62,24){\vector(-2,-3){8}}
\put(88,24){\vector(-2,-1){29}}
\put(39,0){\framebox(20,12){$x_5$}}
 \put(60,2){\pos{l}{$+1$}}
\end{picture}
\caption{\label{fig:regiongraph} The region graph for Forney factor 
graph in~\Fig{fig:rllfactorgraph}.}
\end{figure}

Such a choice for the basic regions seems plausible since the validity of a given
array can be determined by verifying the constraints in each region
and sliding the basic regions along the rows and along the columns of the array.
For a 2-D $(1,\infty)$-RLL constraint, Fig.~\ref{fig:rllfactorgraph} shows
the basic regions and Fig.~\ref{fig:regiongraph} shows the region graph
and the counting numbers associated with each region.


After forming the region graph using the cluster variation method,
we perform GBP on this graph by sending messages 
between the regions while performing exact computations inside each 
region.

We will need the region-based free energy to estimate 
the number of arrays that satisfy a given constraint.
Therefore, we operate GBP on the corresponding
region graph until convergence 
and use the obtained region beliefs 
$\{b_{R}({\bf x}_{R})\}$
to compute the region-based free energy $\hat F_H$ 
(as an estimate of $F_{H}$).
The region-based free energy $\hat F_H$
can then be used to estimate the partition function $Z$ using~\eqref{FHZ}.
We compute $\hat F_H$ as
\begin{eqnarray}
\label{countingproblemfreeenergy}
\hat F_{H} \! \! \! & = & \!\! \! \! \min_{\{b_{R}\}}F_{\mathcal{R}}(\{b_{R}({\bf x}_{R})\})\nonumber \\
\label{numericalissue}
 \! \! \! \! \!  & = & \!\! \! \! \!
\sum_{R\in \mathcal{R}} \!\!\! c_{R} \! \sum_{{\bf x}_{R}}b_{R}({\bf x}_{R})\Big(\!\ln b_{R}({\bf x}_{R}) - \ln \!\!\!\prod_{a\in A_R} \!\! f_a({\bf x}_{a})\!\Big)
\end{eqnarray}

Here $\mathcal{R}$ denotes the set of all regions,
$c_R$ is the counting number, ${\bf x}_R$ stands 
for the set of variables in region $R$, and $A_R$ is the set
of factors in region $R$. See Fig.~\ref{fig:regiongraph}.


\section{Capacity of Noiseless 2-D RLL Constraints}
\label{sec:2Dexample}

For a 2-D RLL constrained channel of width $m$ and of size 
$N = m\times m$, we run GBP on the corresponding region graph to 
compute $\hat F_H$ and an estimate of $Z$. We can then compute
\begin{equation}
\label{Emn}
C(m,m)= \frac{\log_2 Z(m,m)}{m\times m},
\end{equation}
where $Z(m,m)$ denotes the number of 2-D binary arrays of size $m\times m$ 
that satisfy the constraint.

In our numerical experiments in Section~\ref{sec:NumericsC},
for different RLL constraints
we show convergence of $C(m,m)$ to the Shannon capacity 
as $m$ increases.

For example, let us consider a 2-D $(1,\infty)$-RLL constraint
with corresponding Forney factor graph in \Fig{fig:2DGrid}.
For this constraint, we chose basic regions with size 
$2 \times 2$ in 
a sliding window manner over the factor graph, see \Fig{fig:rllfactorgraph}.
Starting from such basic regions, we applied the cluster variation method on 
the factor graph in~\Fig{fig:rllfactorgraph} to obtain the corresponding region 
graph depicted in~\Fig{fig:regiongraph}. The counting numbers $\{c_R\}$ 
are shown next to each region.



\begin{figure}[t]
	\centering
		\includegraphics[width=\linewidth]{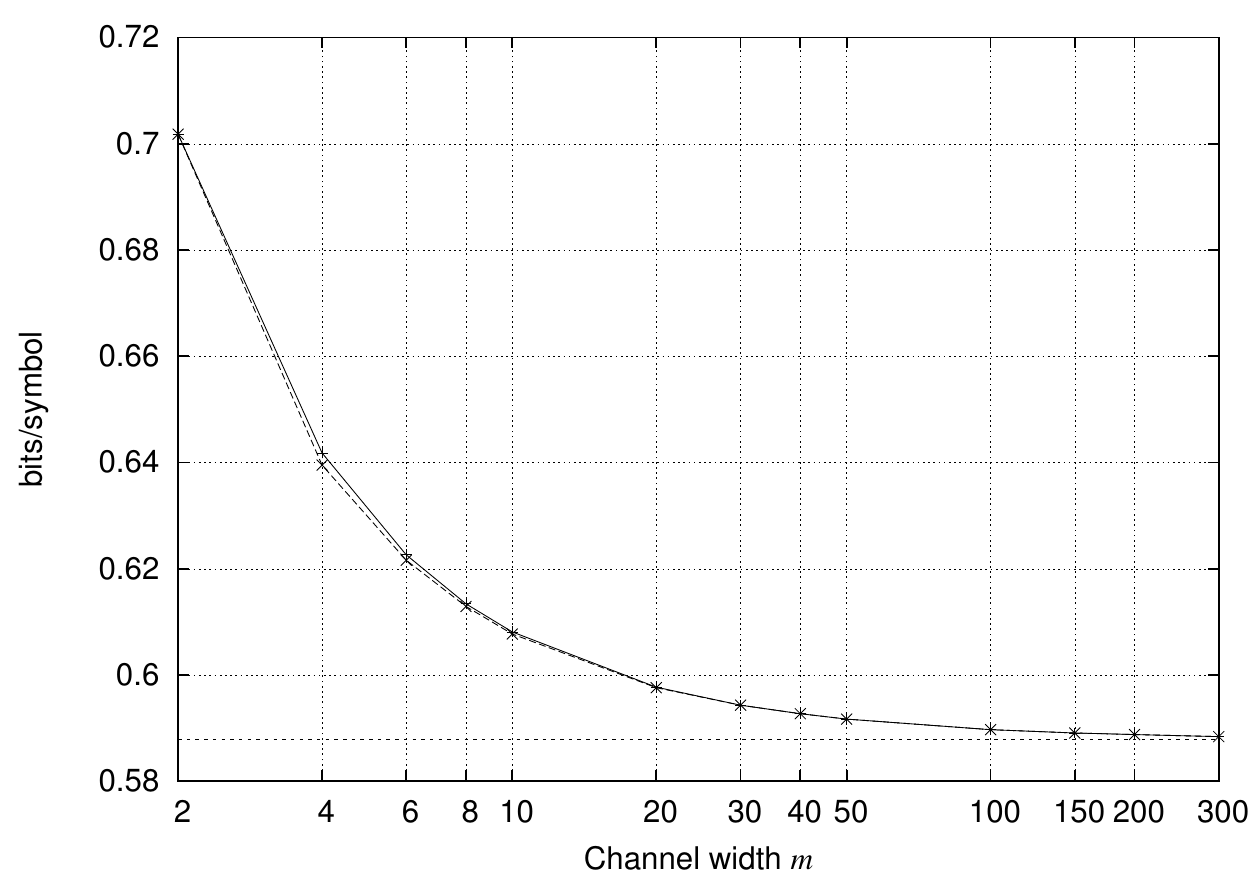}
	\caption{Estimated capacity (in bits per symbol) vs.\ channel width $m$ for a \mbox{2-D $(1, \infty)$-RLL} constraint.
		The horizontal dotted line shows the Shannon capacity for this channel as in~(\ref{cap2D}).}
	\label{fig:capacity11}


	\centering
		\includegraphics[width=\linewidth]{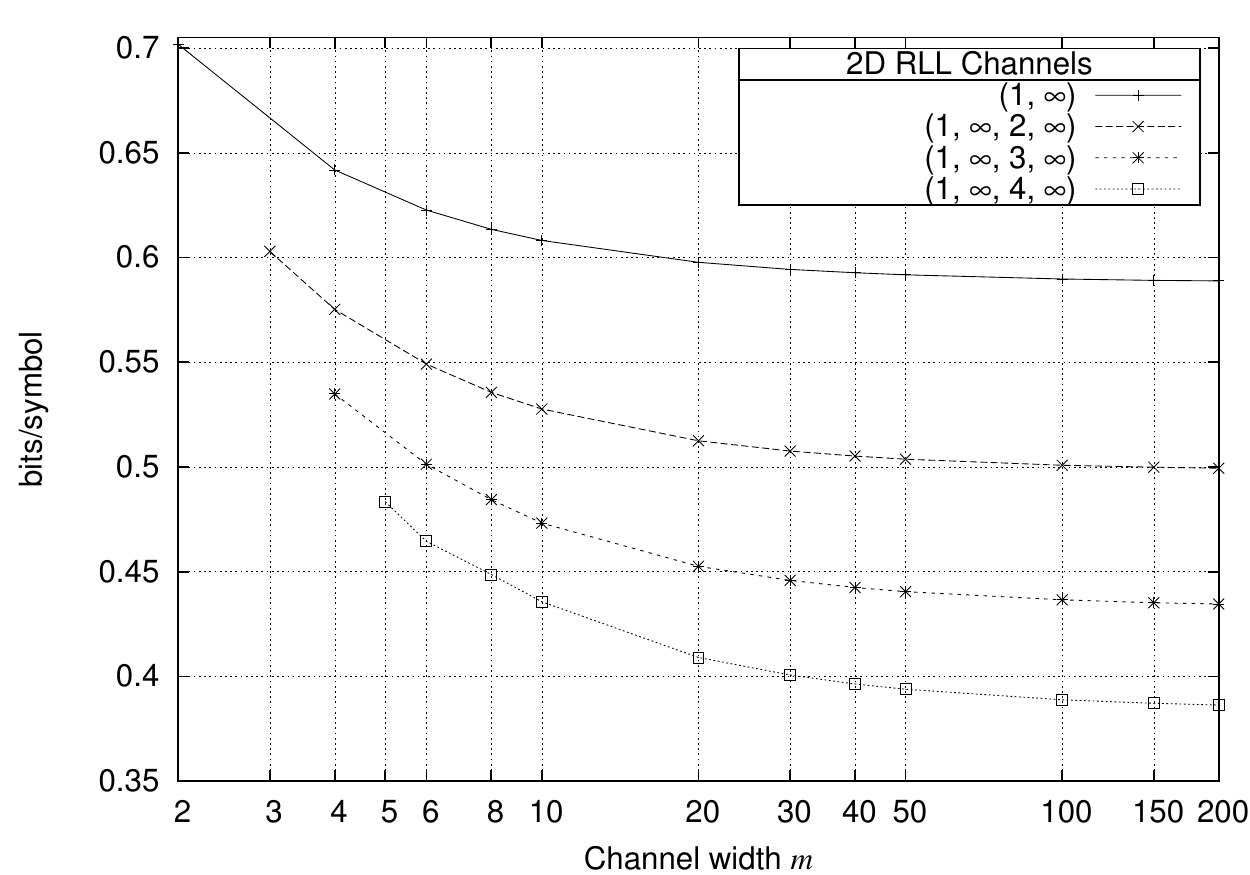}
	\caption{Estimated capacities (in bits per symbol) vs.\ channel width $m$ for a class of \mbox{2-D $(1, \infty, d, \infty)$-RLL} constraints
		 with $d = (1,2,3,4)$.}
	\label{fig:capacitycurves}
\end{figure}


\begin{figure}[t]
	\centering
		\includegraphics[width=\linewidth]{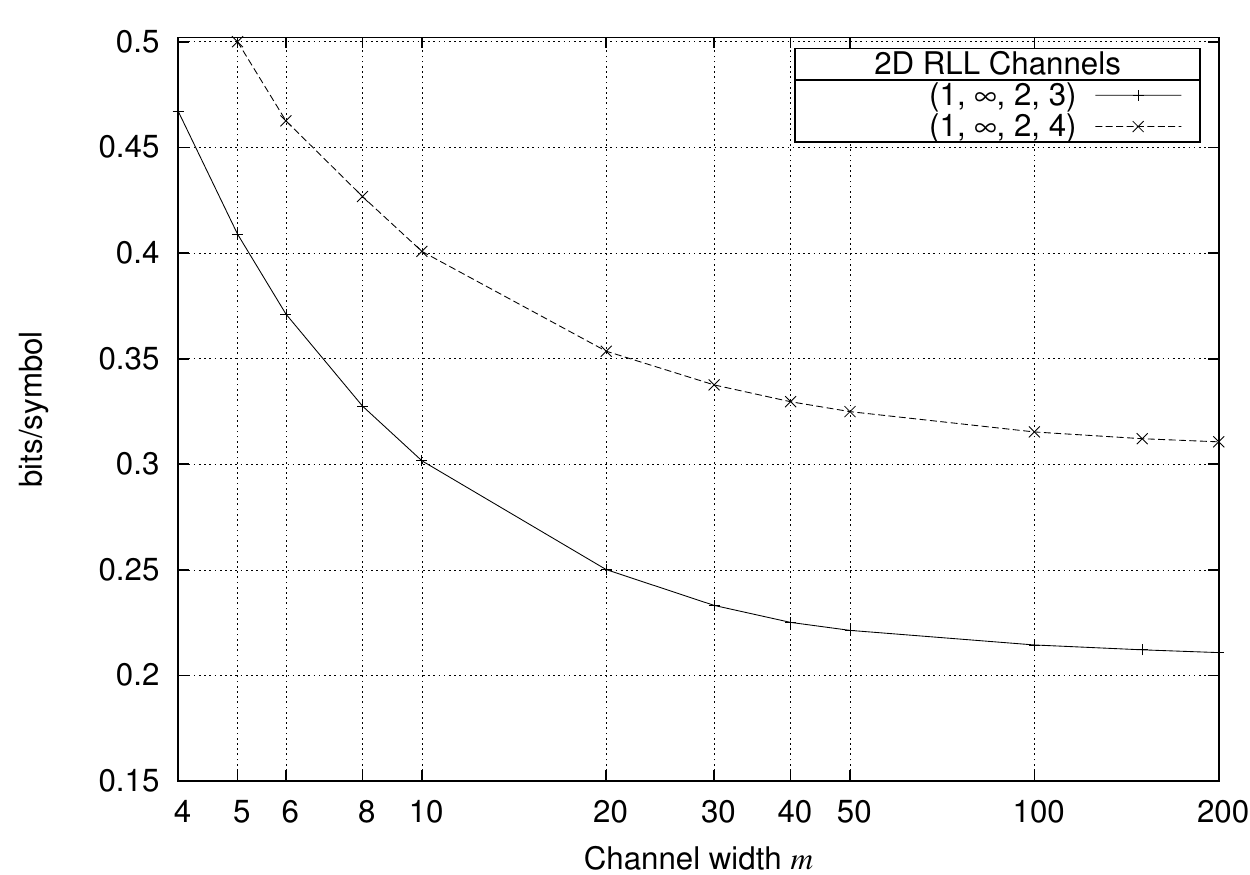}
	\caption{Estimated capacity (in bits per symbol) vs.\ channel width $m$ for a 2-D \mbox{$(1, \infty,2,4)$-RLL} and \mbox{$(1, \infty,2,3)$-RLL} constraints.}
	\vspace{3mm}
	\label{fig:capacitycurves3}
	\centering
		\includegraphics[width=\linewidth]{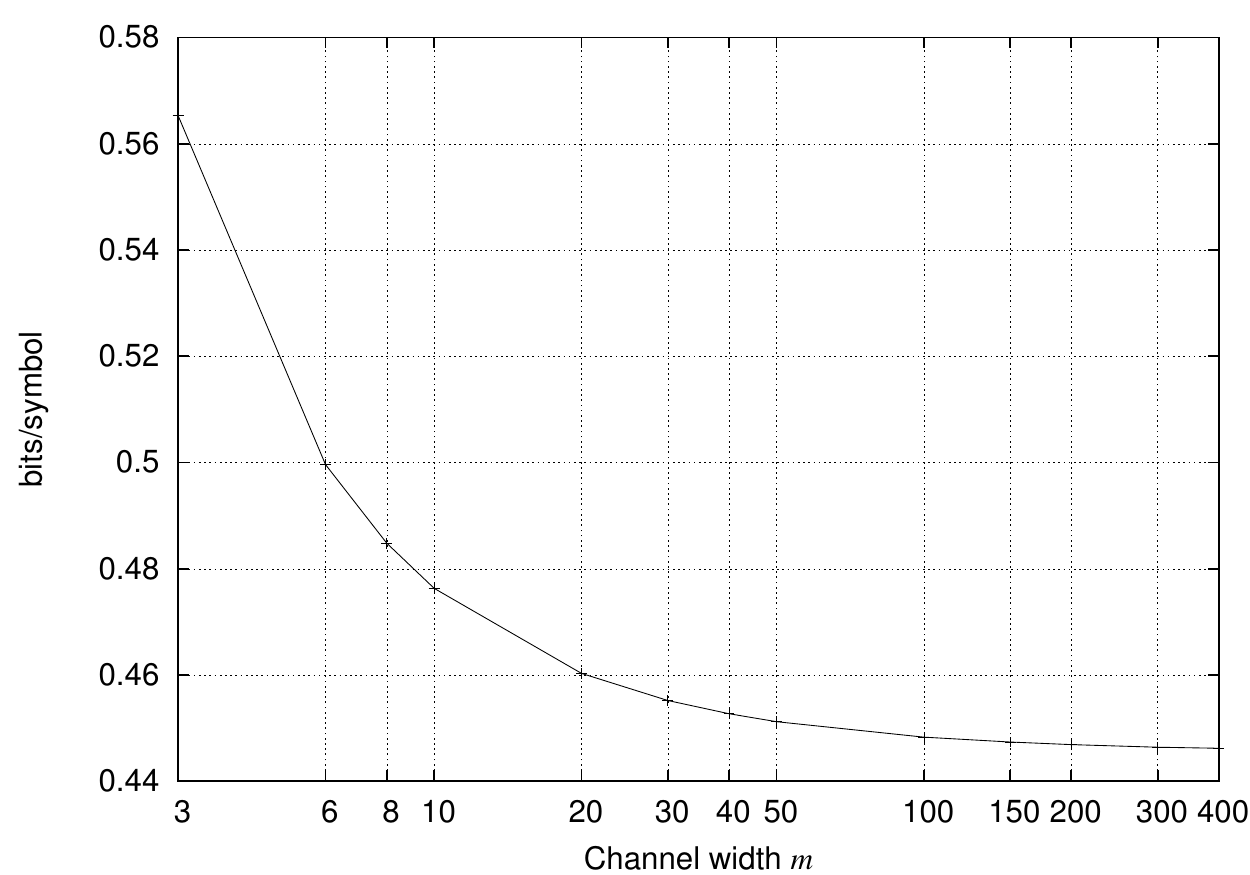}
	\caption{Estimated capacity (in bits per symbol) vs.\ channel width $m$ for a 2-D $(2, \infty)$-RLL constraint.}
	\label{fig:capacitycurves2}

\end{figure}
\vspace{7mm}
\begin{figure}[th!!]
	\centering
		\includegraphics[width=\linewidth]{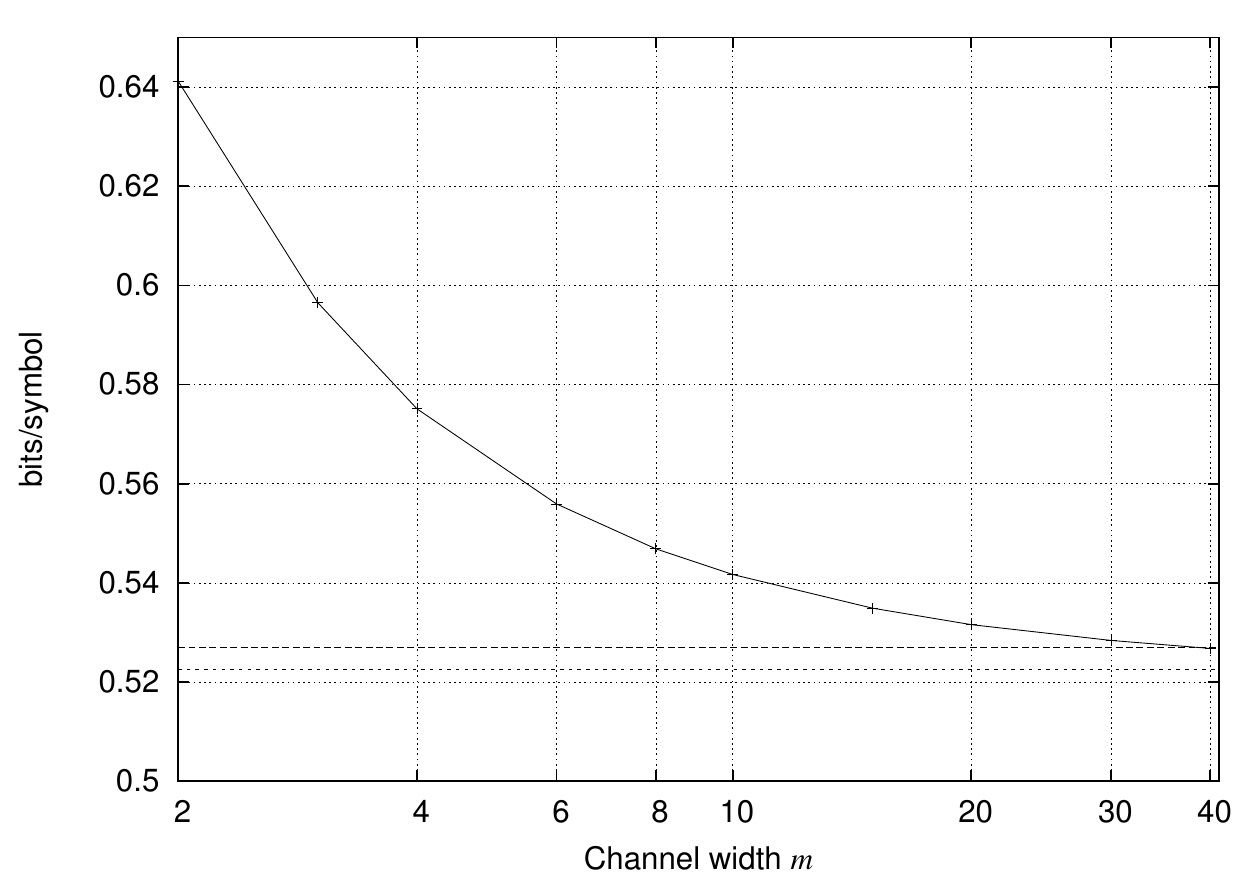}
	\caption{Estimated capacity (in bits per symbol) vs.\ channel width $m$ for a \mbox{3-D $(1, \infty)$-RLL} constraint. The
		horizontal dotted lines show upper and lower bounds on the Shannon capacity for this channel
                as in~(\ref{cap3D}).}
	\label{fig:capacity111}
\end{figure}


\subsection{Numerical Experiments}
\label{sec:NumericsC}

Here we present the numerical results of applying GBP to estimate 
the finite-sized noiseless capacity of RLL constraints.

Tight lower and upper bounds were given for 
the Shannon capacity of a 2-D $(1,\infty)$-RLL constraint in~\cite{CW:98}. 
The bounds were further improved in~\cite{WB:98} and~\cite{NZ:00}, now 
known to nine decimal digits.
\begin{equation}
\label{cap2D}
0.5878911617... \le C_{2D}^{(1,\infty)} \le 0.5878911618...
\end{equation}

For this constraint, 
\Fig{fig:capacity11} shows $C(m,m)$ defined 
in~\eqref{Emn} versus the channel width $m$ over the
interval $[2,300]$.
The estimation was performed using
the parent-to-child and two-way GBP algorithms.
The two algorithms give almost identical results.
The horizontal line in \Fig{fig:capacity11} shows the Shannon 
capacity for this channel in~\eqref{cap2D}.
For a channel of width $300$, the estimated noiseless
capacity is about $0.5884$.

Shown in Fig. \ref{fig:capacitycurves} are plots of $C(m,m)$ 
for 2-D $(1, \infty, d, \infty)$-RLL constraints with 
$d = (1,2,3,4)$ from top to bottom, versus the channel width $m$ over the interval $[2,200]$.
Fig. \ref{fig:capacitycurves3} shows the plots of $C(m,m)$ 
for 2-D $(1, \infty, 2, 4)$-RLL and $(1, \infty, 2, 3)$-RLL constraints
versus $m$ over the interval $[4,200]$.
From our simulation results, for a channel of width $200$ the estimated 
noiseless capacities for 2-D $(1, \infty, d, \infty)$-RLL constraints with $d = (2,3,4)$ are
about $(0.4994, 0.4346,0.3864)$ and the estimated 
noiseless capacities for 2-D $(1, \infty, 2, 4)$-RLL and $(1, \infty, 2, 3)$-RLL
are about  $(0.3106, 0.2109)$.
To the best of our knowledge, no theoretical upper or lower bounds
exist for these constraints.
All plots are obtained using the parent-to-child algorithm.
Note that \mbox{2-D $(1, \infty, 1, \infty)$-RLL} plot in Fig. \ref{fig:capacitycurves}
is the same as the plot in~\Fig{fig:capacity11}.

Also shown in Fig. \ref{fig:capacitycurves2} 
is the plot of $C(m,m)$ for a 2-D $(2, \infty)$-RLL constraint 
versus $m$ over the interval $[3,400]$.
For a channel of 
width $400$, the estimated
noiseless capacity is about $0.4462$.
Best known lower and upper bounds for 
the Shannon capacity of a 2-D $(2,\infty)$-RLL constraint 
are given in~\cite{OR:09} and~\cite{TR:09} respectively, as 
\begin{equation}
\label{cap2D2inf}
0.4453 \le C_{2D}^{(2,\infty)} \le 0.4457
\end{equation}

Our proposed method can be generalized to compute the noiseless 
capacity of 3-D and higher dimensional RLL constraints. 
For a 3-D $(1,\infty)$-RLL constraint the following
lower and upper bounds were introduced in~\cite{NZ:00}
\begin{equation}
\label{cap3D}
0.5225017418... \le C_{3D}^{(1,\infty)} \le 0.5268808478...
\end{equation}

Fig. \ref{fig:capacity111} shows the noiseless capacity 
estimates of a 3-D $(1, \infty)$-RLL constraint, obtained using the parent-to-child algorithm, 
versus the channel width $m$.
The horizontal dotted lines show the upper and lower bounds for 
the Shannon capacity. 
For a channel of width $m = 40$ the GBP estimated capacity is about $0.5267$ 
which falls within these  bounds.

Simulation results and numerical values for the noiseless capacity of many other 2-D RLL
constraints are reported in~\cite{MSthesis}.

\subsection{Bounds for the Shannon Capacity}

For any finite $m$, it is possible to compute lower and upper 
bounds on the Shannon (infinite-size) capacity using $C(m,m)$ the capacity 
of a 2-D RLL constrained channel of width $m$.

For example, consider a 2-D $(1,\infty)$-RLL constraint with local kernels
as in~(\ref{NoAdjacentOnes}). From tiling the whole plane with $m\times m$ squares,
it is clear that $C(m,m)$ is an upper bound for the Shannon capacity
$C_{2D}^{(1,\infty)}$.
On the other hand,
by tiling the plane with $m\times m$ squares 
separated by all-zero guard rows and all-zero guard columns,
we obtain $(\frac{m}{m+1})^2C(m,m) \leq C_{2D}^{(1,\infty)}$.

From Fig. \ref{fig:capacity11}, the estimated capacity at $m=300$ is about
$C(300,300)=0.5884$, we thus obtain the following lower and upper bounds 
for the Shannon capacity
\[
0.5844 \leq C_{2D}^{(1,\infty)} \leq 0.5884.
\]

Note that although GBP performs remarkably well for 2-D constrained 
channels, it is an approximate algorithm which yields approximations
to the lower and upper bounds to the Shannon capacity.
However in order to achieve a desired precision, the bounds could
provide a criterion for
choosing the value of $m$.

\begin{figure}
\setlength{\unitlength}{0.9mm}
\centering

\begin{picture}(87,73)(0,-9)
\small
%
\put(0,60){\framebox(4,4){$=$}}
 \put(4,60){\line(4,-3){4}}
 \put(8,54){\framebox(3,3){}}
 \put(11,54){\line(4,-3){4}}     \put(14,53.5){$y_1$}
\put(4,62){\line(1,0){8}}        \put(8,64){\cent{$x_1$}}
\put(12,60){\framebox(4,4){}}
\put(16,62){\line(1,0){8}}
\put(24,60){\framebox(4,4){$=$}}
 \put(28,60){\line(4,-3){4}}
 \put(32,54){\framebox(3,3){}}
 \put(35,54){\line(4,-3){4}}     \put(38,53.5){$y_2$}
\put(28,62){\line(1,0){8}}       \put(32,64){\cent{$x_2$}}
\put(36,60){\framebox(4,4){}}
\put(40,62){\line(1,0){8}}
\put(48,60){\framebox(4,4){$=$}}
 \put(52,60){\line(4,-3){4}}
 \put(56,54){\framebox(3,3){}}
 \put(59,54){\line(4,-3){4}}     \put(62,53.5){$y_3$}
\put(52,62){\line(1,0){8}}       \put(56,64){\cent{$x_3$}}
\put(60,60){\framebox(4,4){}}
\put(64,62){\line(1,0){8}}
\put(72,60){\framebox(4,4){$=$}}
 \put(76,60){\line(4,-3){4}}
 \put(80,54){\framebox(3,3){}}
 \put(83,54){\line(4,-3){4}}
\put(2,54){\line(0,1){6}}
\put(0,50){\framebox(4,4){}}
\put(2,50){\line(0,-1){6}}
\put(26,54){\line(0,1){6}}
\put(24,50){\framebox(4,4){}}
\put(26,50){\line(0,-1){6}}
\put(50,54){\line(0,1){6}}
\put(48,50){\framebox(4,4){}}
\put(50,50){\line(0,-1){6}}
\put(74,54){\line(0,1){6}}
\put(72,50){\framebox(4,4){}}
\put(74,50){\line(0,-1){6}}
\put(0,40){\framebox(4,4){$=$}}
 \put(4,40){\line(4,-3){4}}
 \put(8,34){\framebox(3,3){}}
 \put(11,34){\line(4,-3){4}}
\put(4,42){\line(1,0){8}}
\put(12,40){\framebox(4,4){}}
\put(16,42){\line(1,0){8}}
\put(24,40){\framebox(4,4){$=$}}
 \put(28,40){\line(4,-3){4}}
 \put(32,34){\framebox(3,3){}}
 \put(35,34){\line(4,-3){4}}
\put(28,42){\line(1,0){8}}
\put(36,40){\framebox(4,4){}}
\put(40,42){\line(1,0){8}}
\put(48,40){\framebox(4,4){$=$}}
 \put(52,40){\line(4,-3){4}}
 \put(56,34){\framebox(3,3){}}
 \put(59,34){\line(4,-3){4}}
\put(52,42){\line(1,0){8}}
\put(60,40){\framebox(4,4){}}
\put(64,42){\line(1,0){8}}
\put(72,40){\framebox(4,4){$=$}}
 \put(76,40){\line(4,-3){4}}
 \put(80,34){\framebox(3,3){}}
 \put(83,34){\line(4,-3){4}}
\put(2,34){\line(0,1){6}}
\put(0,30){\framebox(4,4){}}
\put(2,30){\line(0,-1){6}}
\put(26,34){\line(0,1){6}}
\put(24,30){\framebox(4,4){}}
\put(26,30){\line(0,-1){6}}
\put(50,34){\line(0,1){6}}
\put(48,30){\framebox(4,4){}}
\put(50,30){\line(0,-1){6}}
\put(74,34){\line(0,1){6}}
\put(72,30){\framebox(4,4){}}
\put(74,30){\line(0,-1){6}}
\put(0,20){\framebox(4,4){$=$}}
 \put(4,20){\line(4,-3){4}}
 \put(8,14){\framebox(3,3){}}
 \put(11,14){\line(4,-3){4}}
\put(4,22){\line(1,0){8}}
\put(12,20){\framebox(4,4){}}
\put(16,22){\line(1,0){8}}
\put(24,20){\framebox(4,4){$=$}}
 \put(28,20){\line(4,-3){4}}
 \put(32,14){\framebox(3,3){}}
 \put(35,14){\line(4,-3){4}}
\put(28,22){\line(1,0){8}}
\put(36,20){\framebox(4,4){}}
\put(40,22){\line(1,0){8}}
\put(48,20){\framebox(4,4){$=$}}
 \put(52,20){\line(4,-3){4}}
 \put(56,14){\framebox(3,3){}}
 \put(59,14){\line(4,-3){4}}
\put(52,22){\line(1,0){8}}
\put(60,20){\framebox(4,4){}}
\put(64,22){\line(1,0){8}}
\put(72,20){\framebox(4,4){$=$}}
 \put(76,20){\line(4,-3){4}}
 \put(80,14){\framebox(3,3){}}
 \put(83,14){\line(4,-3){4}}
\put(2,14){\line(0,1){6}}
\put(0,10){\framebox(4,4){}}
\put(2,10){\line(0,-1){6}}
\put(26,14){\line(0,1){6}}
\put(24,10){\framebox(4,4){}}
\put(26,10){\line(0,-1){6}}
\put(50,14){\line(0,1){6}}
\put(48,10){\framebox(4,4){}}
\put(50,10){\line(0,-1){6}}
\put(74,14){\line(0,1){6}}
\put(72,10){\framebox(4,4){}}
\put(74,10){\line(0,-1){6}}
\put(0,0){\framebox(4,4){$=$}}
 \put(4,0){\line(4,-3){4}}
 \put(8,-6){\framebox(3,3){}}
 \put(11,-6){\line(4,-3){4}}
\put(4,2){\line(1,0){8}}
\put(12,0){\framebox(4,4){}}
\put(16,2){\line(1,0){8}}
\put(24,0){\framebox(4,4){$=$}}
 \put(28,0){\line(4,-3){4}}
 \put(32,-6){\framebox(3,3){}}
 \put(35,-6){\line(4,-3){4}}
\put(28,2){\line(1,0){8}}
\put(36,0){\framebox(4,4){}}
\put(40,2){\line(1,0){8}}
\put(48,0){\framebox(4,4){$=$}}
 \put(52,0){\line(4,-3){4}}
 \put(56,-6){\framebox(3,3){}}
 \put(59,-6){\line(4,-3){4}}
\put(52,2){\line(1,0){8}}
\put(60,0){\framebox(4,4){}}
\put(64,2){\line(1,0){8}}
\put(72,0){\framebox(4,4){$=$}}
 \put(76,0){\line(4,-3){4}}
 \put(80,-6){\framebox(3,3){}}
 \put(83,-6){\line(4,-3){4}}
\end{picture}
\caption{\label{fig:SourceChannelGraph}%
Extension of \Fig{fig:2DGrid} to a
Forney factor graph of 
$p(\x,\y)$ 
with $p(\y|\x)$ as in~(\ref{eqn:ChannelIO}).
}
\end{figure}


\section{Information Rates of Noisy 2-D RLL Constraints}
\label{sec:IR}

As explained in Section~\ref{sec:Set-up}, the problem of computing
mutual information rates reduces to computing the output probability.
Therefore, the remaining tasks are
\begin{enumerate} 
\item Drawing input samples $\x^{(1)}, \x^{(2)}, \ldots, \x^{(L)}$ from $\calS_\X$ 
according to $p(\x)$ 
and therefrom creating output samples $\y^{(1)}, \y^{(2)},\ldots, \y^{(L)}$.
\item Computing $p(\y^{(\ell)})$ for each $\ell = 1,2, \ldots, L$.
\end{enumerate}

We will compute $p(\y^{(\ell)})$ based on
\begin{equation}
p(\y^{(\ell)}) = \sum_{\x \in \calS_\X} p(\x)p(\y^{(\ell)}|\x),  \label{eqn:PY2}
\end{equation}
where $p(\x)$ is a probability mass function on $\calS_\X$.

Let us assume uniform distribution over the admissible channel input
configurations. Therefore we have
\begin{eqnarray}
p(\x) &=& |\calS_\X|^{-1} \\ \label{eqn:PXS1}
      &=& 2^{-NC_{2D}}, \label{eqn:PXS2}
\end{eqnarray}

we also assume the channel is memoryless and $p(\y|\x)$ factors as
\begin{equation} \label{eqn:ChannelIO}
p(\y|\x) = \prod_{i=1}^N p(y_i|x_i).
\end{equation}

For such a noisy 2-D constrained channel, the corresponding Forney factor graph, 
as an extension of \Fig{fig:2DGrid}, is shown in \Fig{fig:SourceChannelGraph}.

Using~(\ref{eqn:PXS2}) and~(\ref{eqn:ChannelIO}), we can rewrite (\ref{eqn:PY2}) as
\begin{eqnarray}
p(\y^{(\ell)}) & = & 2^{-NC_{2D}} \sum_{\x \in \calS_\X}\prod_{i=1}^N p(y^{(\ell)}_i|x_i) ,\\  \label{eqn:PY3}
	       & = & 2^{-NC_{2D}}Z(\y^{(\ell)}), \label{eqn:PY4}
\end{eqnarray}
where $Z(\y^{(\ell)})$ has the same form as the sum in~\eqref{Z2}.

The input samples $\x^{(1)}, \x^{(2)}, \ldots, \x^{(L)}$ 
are generated as follows.
We run GBP on 
Fig.~\ref{fig:regiongraph} until convergence to compute 
the region beliefs $\{b_{R}(\x_{R})\}$ at each region $R$.
The region beliefs are GBP approximations to the corresponding 
marginals $\{p_R(\x_R)\}$. In our numerical experiments,
each sample $\x^{(\ell)}$ is then generated 
piecewise sequentially according to 
the beliefs $b_{R}(\x_{R})$ in basic regions.
For example, in the region graph of~\Fig{fig:regiongraph},
after computing $b_R(x_1,x_2,x_4,x_5)$, sample $x_1$ is 
drawn according to $b_R(x_1)$, sample $x_2$ is drawn
according to $b_R(x_2|x_1)$, etc.
The input samples $\x^{(1)}, \x^{(2)}, \ldots, \x^{(L)}$ are then used to create output 
$\y^{(1)}, \y^{(2)},\ldots, \y^{(L)}$ using~(\ref{eqn:ChannelIO}).

The beliefs are directly proportional
to the factor nodes involved in each region, 
which
guarantees that the samples are drawn from $\calS_\X$.
Moreover, since beliefs are  
good approximations to the marginal probabilities, 
one expects that the samples are drawn from a 
distribution close to 
$p(\x)$,
see~\cite{YFW:05}.

In order to compute
$Z(\y^{(\ell)})$,
as in Section~\ref{sec:GBP}, we start from the factor graph in 
\Fig{fig:SourceChannelGraph} to build the region graph representing 
the problem and run GBP on this region graph.
Finally, the estimated $p(\y^{(1)}), p(\y^{(2)}), \ldots, p(\y^{(L)})$ are 
used to compute an estimate of $H(\Y)$ as in~(\ref{HYE}).

\subsection{Numerical Experiments}
\label{sec:NumericsI}

In our numerical experiments we consider \mbox{$(1,\infty)$-RLL} and 
\mbox{$(2,\infty)$-RLL} constrained channels with size 
$N= 30\times 30$ 
and input alphabet 
$\calX = \{-1,+1\}^N$.

Noise is assumed to be i.i.d.\  zero mean Gaussian 
with variance $\sigma^2$ and independent of the input.
We thus have
\begin{equation}
H(\Y|\X) =  \frac{N}{2}\log(2\pi e\sigma^2),
\end{equation}

and $p(\y|\x)$ in~(\ref{eqn:ChannelIO}) has kernels of the form
\begin{equation}
p(y_i|x_i) = \frac{1}{\sqrt{2\pi\sigma^2}}
          \exp\!\bigg(-\frac{1}{2\sigma^2} 
          \Big(y_i - x_i\Big)^2 \bigg).
\end{equation}

SNR is defined as
\begin{equation}
\text{SNR} \eqdef 10\log_{10} \frac{1}{\sigma^2}
\end{equation}

\begin{figure}[t]
\centering
\includegraphics[width = \linewidth]{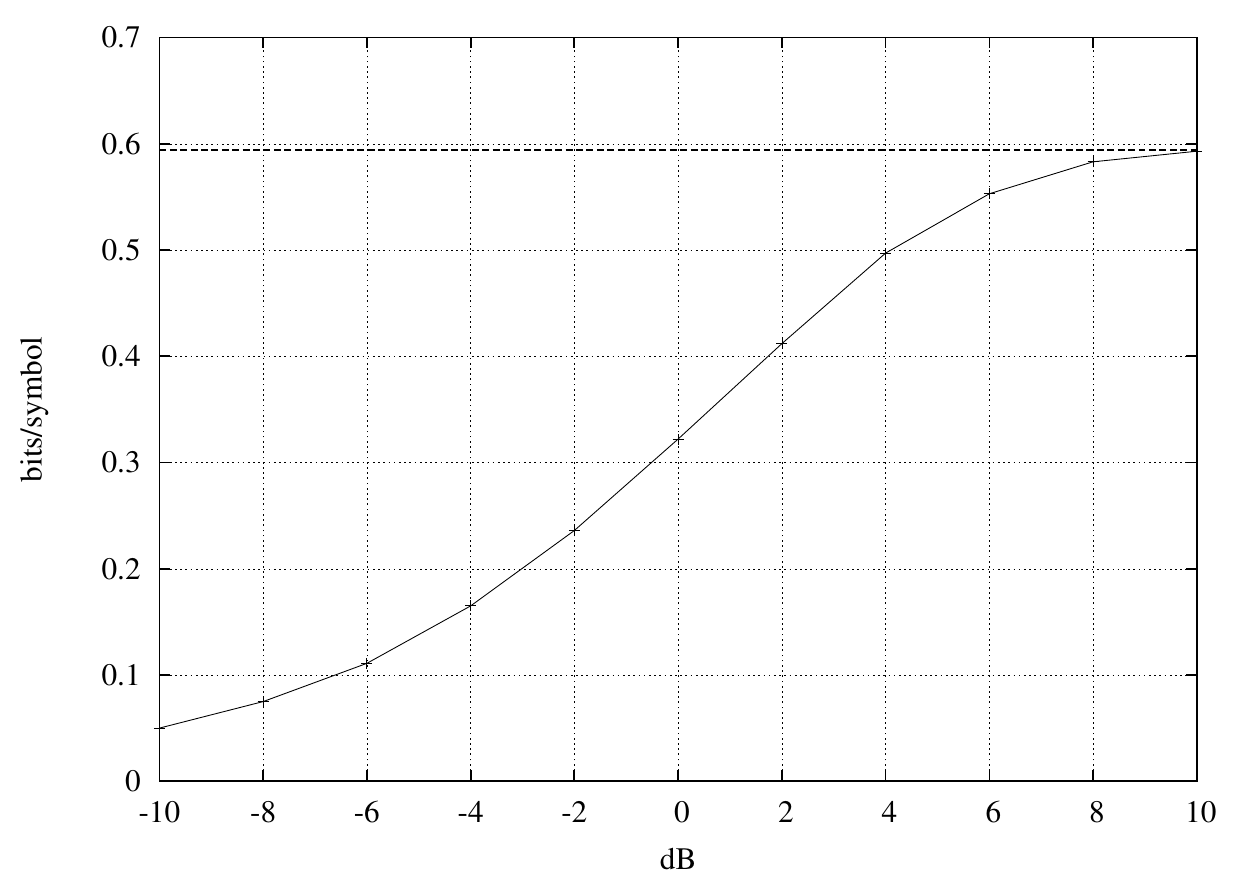}
\caption{\label{fig:IR1inf}%
Estimated information rate (in bits per symbol) vs.\ SNR (in dB) 
for a $30\times 30$ channel with a $(1,\infty)$-RLL constraint and additive
white Gaussian noise.}
\centering
\includegraphics[width= \linewidth]{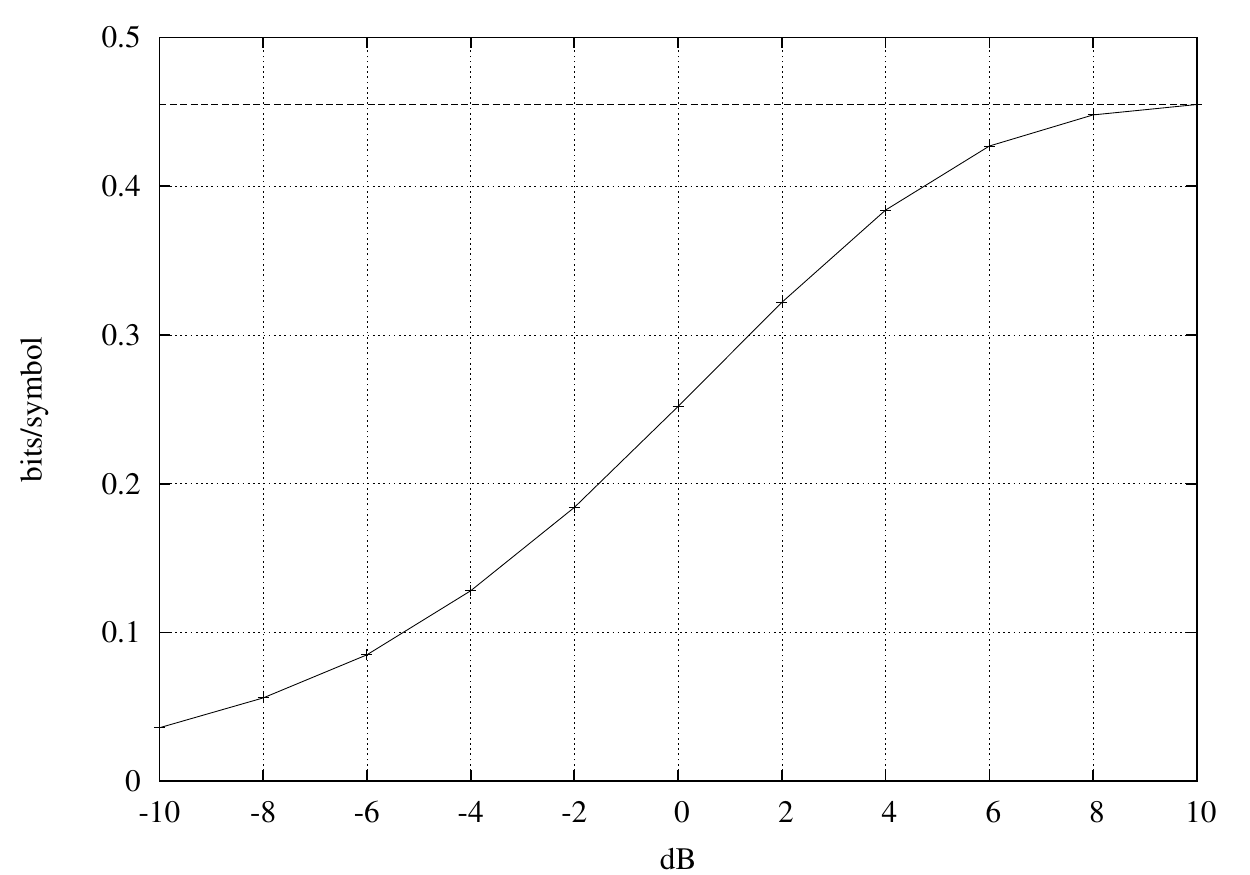}
\caption{\label{fig:IR2inf}%
Estimated information rate (in bits per symbol) vs.\ SNR (in dB) 
for a $30\times 30$ channel with a $(2,\infty)$-RLL constraint and additive
white Gaussian noise.}
\end{figure}

Shown in Fig. \ref{fig:IR1inf} is the estimated information rate 
vs.\ SNR over the interval $[-10,10]$ dB for a noisy 
2-D \mbox{$(1, \infty)$-RLL} constraint. The horizontal dotted line
shows the estimated noiseless capacity which can be read from 
Fig. \ref{fig:capacity11} and is about $0.5943$ for this size of channel.

Illustrated in Fig. \ref{fig:IR2inf} is the estimated information rate
vs.\ SNR over the interval $[-10,10]$ dB
for a noisy 2-D \mbox{$(2, \infty)$-RLL} channel.
The horizontal dotted line
shows the estimated noiseless capacity which can be read 
from Fig. \ref{fig:capacitycurves2}
and is about $0.4552$ for this size of channel.

The simulation results were obtained by averaging over \mbox{$L = 1000$}
realizations of the channel output.

Simulation results and numerical values for mutual information rates of 
many other 2-D RLL constraints are reported in~\cite{MSthesis}.

\section{Concluding remarks}

We proposed a GBP-based method 
to estimate the noiseless capacity and mutual information rates of 
RLL constraints in two and three dimensions.
For noiseless RLL constraints, the method
was applied to estimate the finite-size capacity of different 
constraints and to show convergence to the Shannon capacity as the 
size of the channel increases. In particular, the proposed method can 
be used to estimate the noiseless capacity of RLL constraints in the 
cases that the capacity is not known to a useful accuracy.
The method was also applied to estimate mutual information rates of
noisy RLL constraints with additive white Gaussian noise and with a 
uniform distribution over the admissible input configurations.
Our simulation results show mutual information rates of different constraints
as a function of SNR.



\newcommand{\IT}{IEEE Trans.\ Inform.\ Theory}
\newcommand{\CASI}{IEEE Trans.\ Circuits \& Systems~I}
\newcommand{\COM}{IEEE Trans.\ Comm.}
\newcommand{\COMLet}{IEEE Commun.\ Lett.}
\newcommand{\COMMag}{IEEE Communications Mag.}
\newcommand{\ETT}{Europ.\ Trans.\ Telecomm.}
\newcommand{\SPMag}{IEEE Signal Proc.\ Mag.}
\newcommand{\ProcIEEE}{Proceedings of the IEEE}

\section*{Acknowledgements}

The authors gratefully acknowledge the support of Prof.\ \mbox{H.-A.}~Loeliger.
The first author wishes to thank Ori Shental for his helpful comments 
on GBP implementation. We would also like to thank the reviewers for their
many helpful suggestions
that helped to improve the presentation of our paper.

\newpage







\end{document}